\documentclass[12pt]{article}

\usepackage{epsfig}
\usepackage{amsmath}
\usepackage{amssymb}

\begin{document}

\title{Time scales for fission at finite temperature}
\author{H. Hofmann $^{1}$, F.A. Ivanyuk$^{1,2}$\\ \small\it{1)
Physik-Department der TU M\"unchen, D-85747 Garching, Germany } \\
\small\it{2) Institute for Nuclear Research, 03028 Kiev-28,
Ukraine}}
\date{September 30, 2002}
 \maketitle

\vspace{-0.5cm}
\begin{center}
\large{"Symposium on Nuclear Clusters", \\ Rauischholzhausen,
Germany, 5-9 August 2002}
\end{center}
\vspace{0.5cm}
\begin{abstract} The concept of the "transient effect" is examined in
respect of a "mean first passage time". It is demonstrated that
the time the fissioning system stays inside the barrier is much
larger than suggested by the transient time, and that no
enhancement of emission of neutrons over that given by Kramers'
rate formula ought to be considered.

\end{abstract}

\section{Introduction}\label{intro}

Fission at finite thermal excitation is characterized by the
evaporation of light particles and $\gamma$'s. Any description of
such a process must rely on statistical concepts, both with
respect to fission itself as well as with respect to particle
emission. For decades it has been customary to describe
experiments in terms of particle\footnote{To simplify the
discussion we will not distinguish the nature of the "particles"
and in this sense include $\gamma$'s in this notation.} and
fission widths, where the former, $\Gamma_n$, is identified
through the evaporation rate and the latter $\Gamma_f$ is given by
the Bohr-Wheeler formula $\Gamma_f\equiv \Gamma_{\rm{BW}}$ for the
fission rate. Often in the literature this is referred to as the
"statistical model". It was only in the 80's that discrepancies of
this procedure with experimental evidence was encountered: Sizably
more neutrons were seen to accompany fission events than given by
the ratio $\Gamma_n/\Gamma_{\rm{BW}}$ (for a review see e.g.
\cite{pauthoe}). It was then that one recalled Kramers' old
objection \cite{kram} against the Bohr-Wheeler formula. Indeed, in
this seminal paper he pointed to the deficiency of the picture of
Bohr and Wheeler in that it discards the influence of couplings of
the fission mode to the nucleonic degrees of freedom. Such
couplings will in general reduce the flux across the barrier as
the energy in the fission degree of freedom $Q$ may be diminished
and fall below the barrier.

In Kramers' picture this effect is realized through the presence
of frictional and fluctuating forces (intimately connected to each
other by the fluctuation dissipation theorem). For not too weak
dissipation Kramers' rate formula
writes\begin{equation}\label{kram-rate} \Gamma_{\rm{K}}=
\frac{\hbar\varpi_{\rm{a}} }{ 2\pi}
\exp\left(-\frac{E_{\rm{b}}}{T}\right)\,
\left(\sqrt{1+\eta_{\rm{b}}^2} - \eta_{\rm{b}}\right)
=\frac{\hbar}{\tau_{\rm{K}}}=\hbar R_{\rm{K}} .\end{equation}
Here, $T$ and $E_{\rm{b}}$ stand for temperature and barrier
height, $\varpi_{\rm{a}}$ for the frequency of the motion in the
(only) minimum at $Q=Q_{\rm{a}}$ and $\eta_{\rm{b}}=(\gamma/(2M
\varpi))_{\rm{b}}$ for the dissipation strength at the barrier (at
$Q=Q_{\rm{b}}$) with $\gamma$ being the friction coefficient and
$M$ the inertia. For the sake of simplicity we will assume these
coefficients not to vary along the fission path; otherwise the
formula must be modified \cite{hiry}. For vanishing dissipation
strength (\ref{kram-rate}) reduces to the Bohr-Wheeler formula
(simplified to the case that the equilibrium of the nucleons can
be parameterized by a temperature).

Commonly, formula (\ref{kram-rate}) is derived in a time dependent
picture solving the underlying Fokker-Planck equation for special
initial conditions with respect to the time dependence of the
distribution function\footnote{For Kramers' equation proper this
involves the coordinate $Q$ and a momentum $P$. The latter is
absent in the Smoluchoski equation into which the former turns
into for overdamped motion.}. The initial condition is intimately
related to the condition of a compound reaction, in that the decay
process is assumed to be independent of how the compound nucleus
is produced. In this spirit one does not need to care for the
truly initial stages of the reaction. One simply assumes the
system to be located initially around the ground state minimum
$Q=Q_{\rm{a}}$ of the static energy. However, there still is room
for the precise definition, as the system may or may not be in
(quasi-)equilibrium with respect to the fission degree of freedom
$Q$ --- which for large damping belongs to the slowest modes
present. Since the corresponding momentum may safely be assumed to
relax much faster one may also let the system start sharply at
$Q=Q_{\rm{a}}$ with a Maxwell distribution in $P$. In any case,
for the current across the barrier it takes some finite time to
build up. This apparent {\em delay} of fission was taken
\cite{graliwei,bhagran} as an indication for {\em additional}
possibilities to emit light particles {\em beyond} the measure
given by $\Gamma_n/\Gamma_{\rm{K}} \,> \,
\Gamma_n/\Gamma_{\rm{BW}}$.

In case that the full process is studied in a time dependent
picture both for fission as well as for particle emission, as done
in the typical Langevin codes \cite{mactheo}, such an effect is
included automatically. Problems arise, however, if one tries to
imitate this delay in statistical codes which are in use for
analyzing experimental results. Such codes are based on {\em time
independent} reaction theory. For this reason it is not obvious
how this method may be reconciled with the picture of fission
delay, the "transient effect". In the present note we like to shed
some light on this problem by exploiting the concept of a "mean
first passage time" (MFPT).

\section{Time dependent current across the barrier}\label{cur-crobar}

\begin{figure} [htb]
\begin{center} \epsfig{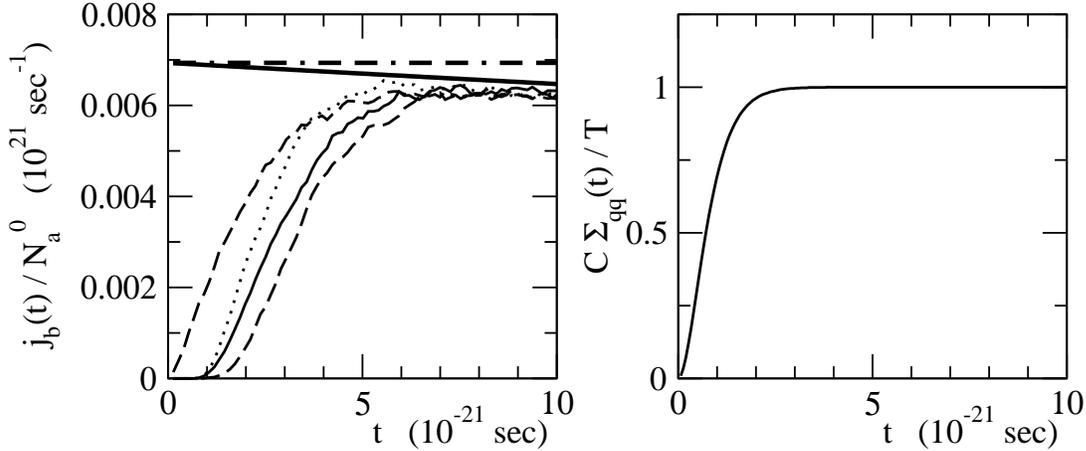} \caption{The
current across the barrier for different initial distributions,
see text.} \label{raufig1}
\end{center}
\end{figure}

Solutions of the transport equation require appropriate boundary
conditions for the coordinate $Q$ (and $P$ if present). For the
solutions discussed above the boundary conditions are chosen to
make sure that the distribution vanishes at infinity. Calculations
of the current  $j(t)$ across the barrier then typically imply a
behavior as exhibited in Fig.\ref{raufig1} for different initial
conditions. In all cases the asymptotic value of $j_{\rm{b}}(t)$
is seen to behave like $\Gamma_{\rm{K}} \exp(-\Gamma_{\rm{K}}
\,t/\hbar)$. The differences at short times are due to the
following choices:

(i) For the dashed curves the system starts out of equilibrium for
both $Q$ and $P$; the curve on the left corresponds to the current
at the barrier $j_{\rm{b}}(t) = j(Q_{\rm{b}},t)$ and the right one
to that in the scission region $j_{\rm{sc}}(t) =
j(Q_{\rm{sc}},t)$. The equilibrium is defined by the oscillator
potential which approximates the $V(Q)$ around $Q_{\rm{a}}$.

(ii) For the fully drawn line the system starts at $Q_{\rm{a}}$
sharp with a Maxwell distribution in $P$. The obvious delay is
essentially due to the relaxation of $Q$ to the quasi-equilibrium
in the well. This feature is demonstrated on the right by the
$t$-dependence of the width in $Q$ (as given by the average
potential energy).

(iii) The dotted curve corresponds to an initial distribution
which is also sharp in the momentum but centers at the finite
value $\langle P\rangle_ {t=0}=\sqrt{MT}$.

%%%%%%%%%%
\begin{figure}%[htb]
\begin{center}
\epsfig{figure= 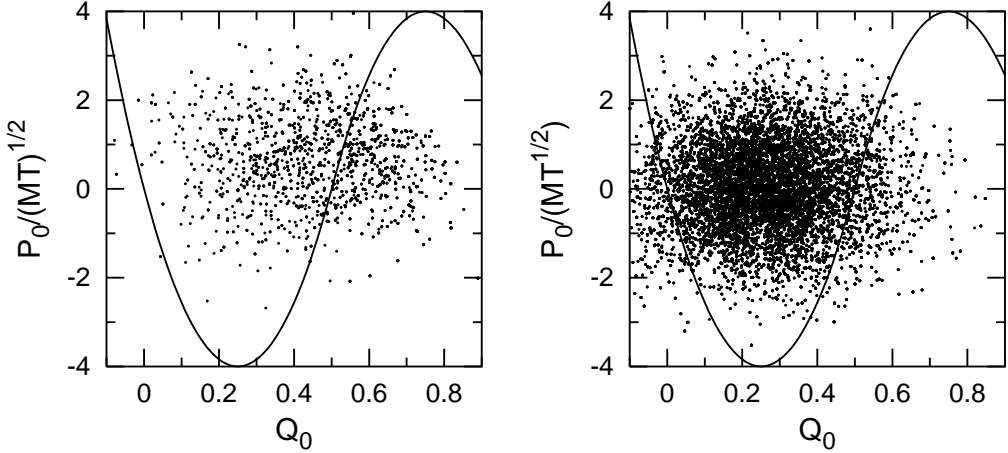, height=6cm}%, width=9cm}% angle=}
\caption{Samples of initial points which overcome the saddle
within a given time $\tau_s$: Left part: $\tau_s\simeq
\tau_{\rm{trans}}$; right part $\tau_s\simeq \tau_{\rm{K}} $.}
\label{raufig2}\vspace*{-0.5cm}
\end{center}
\end{figure}

\vspace*{-0.5cm}
%%%%%%%%%%%%%%%%%
The figure demonstrates clearly a remarkable uncertainty in the
very concept of the "transient effect", namely that the
$j_{\rm{b}}(t)$ reaches its asymptotic behavior only after some
finite time $\tau_{\rm{trans}}$. Even more important appears to be
that the whole effect is due to the arbitrariness in choosing time
zero. If the calculation were repeated at some later time $t_0
>\tau_{\rm{trans}}$, the same behavior would be seen! This is due
to the fact that the whole transient effect only comes about
because in the initial distribution there is some favorable region
in phase space from which it is easiest to reach the barrier. This
is demonstrated in Fig.\ref{raufig2}. There, all initial points in
phase space are sampled which cross the barrier after some time
$\tau_s$. On the right a sufficiently large time was chosen such
that {\em most parts} of the initial distribution have
"fissioned". As exhibited on the left, for the much shorter time,
of the order of $\tau_{\rm{trans}}$, only a small fraction of
points have succeeded to do this, namely those which start close
to the barrier with a more favorite initial momentum. The vast
majority of particles are still waiting to complete the same
motion but at a later time! This feature is very important for
several reasons, in particular (i) for an understanding of the
essentials of the concept of the MFPT
\cite{vankampen,gardiner-STM,Risken}, (ii) that there still is
ample time for neutrons to be emitted inside the barrier, even for
$t \gg \tau_{\rm{trans}}$.
%%%%%%%%%%
%%%%%%%%%%%%%%%%%

The calculations have been performed by simulating the Langevin
equations exploiting a locally harmonic oscillation, for the
following parameters: $T=3$ MeV, $E_b=8$ MeV, $\hbar\varpi_a=1$
MeV and $\eta_a=1$. The potential was constructed from two
oscillators, an upright one and one upside down, joined at some
point between the minimum and the saddle with a smooth first
derivative.

\section{The mean first passage time}\label{conc-MFPT}

For the sake of simplicity we take the example of overdamped
motion for which the momentum is always in equilibrium such that
one only needs to consider the time evolution of the fission
coordinate $Q$. The first passage time may then be defined in the
following way. Suppose that at $t=0$ the particle starts at the
potential minimum $Q_{\rm{a}}$ sharp. Because of the fluctuating
force there will be many trajectories $i$ which will pass a
certain exit point $Q_{\rm{ex}}$ once. This process may take the
time $t_i$, the {\em first passage time}. The {\em mean}-FPT
$\tau_{\rm{mfpt}} (Q_{\rm{a}} \to Q_{\rm{ex}})$ is defined by the
average $\langle t_i \rangle $ over all possibilities. In order to
really obtain the mean {\em first} passage time the $i$ has to be
removed from the ensemble once it has exited the interval at
$Q_{\rm{ex}}$: the "particle" can be said to be absorbed at
$Q_{\rm{ex}}$ ("absorbing barrier"). As the potential $V(Q)$ is
assumed to rise to infinity for $Q\to -\infty$, any motion to the
far left will bounce back: the region $Q\to -\infty$ acts as a
"reflecting barrier".

The MFPT $\tau_{\rm{mfpt}}$ can also be calculated from solutions
$K(Q,t\,|\,Q_{\rm{a}},0)$ of the Smoluchowski equation, adequate
for overdamped motion \cite{vankampen,gardiner-STM,Risken}. The
initial condition for the particles to start at $Q_{\rm{a}}$ is
then: $\lim_{t\to 0 }K(Q,t|Q_{\rm{a}},0)=\delta(Q-Q_{\rm{a}})$,
which is identical to the one used for the fully drawn line of
Fig.\ref{raufig1}. It is this case which (for constant friction
and temperature) allows for an analytic form for the MFPT,
\begin{equation}\label{MFPT-smo} \tau_{\rm{mfpt}}(Q_{\rm{a}} \to Q_{\rm{ex}})=
\frac{\gamma}{T}\int_{Q_{\rm{a}}}^{Q_{\rm{ex}}} du
\;\exp\left[\frac{V(u)}{T}\right] \int_{-\infty}^u dv
\;\exp\left[-\frac{V(v)}{T}\right]\,,\end{equation} by reasoning
as follows: The probability of finding at time $t$ the particle
still inside the interval $(-\infty,~Q_{\rm{ex}})$ is given by
$W(Q_{\rm{a}},t)=\int_{-\infty}^{Q_{\rm{ex}}}\,
 dQ\,K(Q,t\,|\,Q_{\rm{a}},0) $. Hence, the probability for it
to leave the region during the time lap from $t$ to $t+dt$ is
determined by $-d W = -\frac{\partial}{\partial t} W(Q_{\rm{a}},t)
dt$, such that one has $\tau_{\rm{mfpt}}(Q_{\rm{a}}\to
Q_{\rm{ex}})= -\int t d W$ which turns into
\begin{equation}\label{MFPT-prop}\tau_{\rm{mfpt}}(Q_{\rm{a}}\to Q_{\rm{ex}}) =
\int_0^\infty
\int_{-\infty}^{Q_{\rm{ex}}}\,K(Q,t\,|\,Q_{\rm{a}},0) \, dQ \,dt =
\int_0^\infty dt\,t\, j(Q_{\rm{ex}},t\,|\,Q_{\rm{a}},0) \,.
\end{equation} These formulas are associated to  the special
boundary conditions with respect to the
%%%%%%%%%%%%%%%%%
%\setlength{\textfloatsep}{10pt}
\begin{figure}
\begin{center}
\epsfig{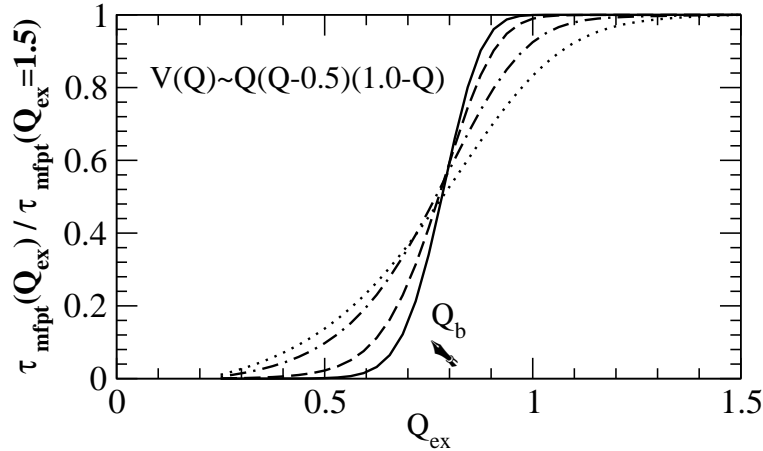}%, width=9cm}% angle=}
\caption{MFPT for the cubic potential normalized to its asymptotic
value. Solid, dashed, dotted-dashed and dotted curves correspond
to $T/E_b=0.1, 0.2, 0.5, 1.0$.} \label{raufig3}\vspace*{-0.5cm}
\end{center}
\end{figure}

%%%%%%%%%%%%%%%%%
coordinate mentioned before, the reflecting barrier at $Q\to
-\infty$ and an absorbing barrier at $Q_{\rm{ex}}$. In particular
for the latter reason it is not permitted to use in
(\ref{MFPT-prop}) the currents shown in Fig.\ref{raufig1}.
Inserting them blindly would indeed lead to expressions for
$\tau_{\rm{mfpt}}$ in which the $\tau_{\rm{trans}}$
appears\footnote{Approximating the form of the current by
$j_{\rm{b}}(t) = j_{\rm{norm}}
(1-e^{-\Gamma_{\rm{trans}}t/\hbar})\exp\left(-\Gamma_{\rm{K}}t/\hbar\right)$
one would falsely get $\tau_{\rm{mfpt}}(Q_{\rm{b}}) =
\hbar/(\Gamma_{\rm{K}}+\Gamma_{\rm{trans}}) +
\hbar/\Gamma_{\rm{K}}$.}. {\em This is in clear distinction to the
correct expression (\ref{MFPT-smo}).} Actually, the derivation of
(\ref{MFPT-smo}) involves proper solutions of that equation which
is "adjoint" to the Smoluchowski equation, and which describes
motion backward in time. In Fig.\ref{raufig3} we show the
dependence of $\tau_{\rm{mfpt}}(Q_{\rm{a}}\to Q_{\rm{ex}})$ on
$Q_{\rm{ex}}$ as given by (\ref{MFPT-smo}) calculated for a cubic
potential. Evidently, the MFPT needed to reach the saddle  at
$Q_{\rm{b}}$ is {\em exactly half the asymptotic value}. The
latter may be identified as the mean fission life time $\tau_f
\equiv \tau_{\rm{mfpt}}(Q_{\rm{a}}\to Q_{\rm{ex}} \gg
Q_{\rm{b}})$. For the typical conditions under which Kramers' rate
formula (\ref{kram-rate}) is valid for overdamped motion, this
relation of $R_{\rm{K}}=\left(\tau_f\right)^{-1}$ to the
asymptotic MFPT can be derived analytically \cite{gardiner-STM}.
Another remarkable feature seen in Fig.\ref{raufig3} is the
insensitivity of the MFPT to the exit point for small and large
$Q_{\rm{ex}}$.

\section{Discussion}\label{discu}

It should be evident from the previous discussion that in the very
concept of the MFPT there is no room for a transient effect. After
all, formula (\ref{MFPT-smo}) is based on {\em exact solutions of
the transport equation which satisfy the same initial condition as
those used for the plots in Fig.\ref{raufig1}.} One essential
difference is seen in the fact that the evaluation of the MFPT
takes into account an average over {\em all} initial points, as is
warranted by the definition of the MFPT through the probability
distribution $-dW$. Contrasting this feature, and as outlined in
sect.\ref{cur-crobar}, the transient effect only represents a
minor part of the initial population, namely that one which
reaches the barrier first. Discarding the rest implies ignoring
the many particles which are still moving inside the barrier for
times typically much longer than $\tau_{\rm{mfpt}}(Q_{\rm{a}}\to
Q_{\rm{b}})$. Neutrons from deformations corresponding to that
region are not only emitted within $\tau_{\rm{trans}}$ but within
$\tau_{\rm{mfpt}}(Q_{\rm{a}}\to Q_{\rm{b}})$, which turns out to
be just half of the total fission time $\tau_{\rm{K}}$. Of course,
this discussion shows that --- besides the additional neutrons
often associated to the transient effect --- also those,
supposedly emitted during the motion from saddle to scission are
not treated correctly by introducing the saddle to scission time
$\tau_{\rm{ssc}}$ of \cite{hofnix}. As one may guess from
Fig.\ref{raufig3}, like the $\tau_{\rm{trans}}$, the
$\tau_{\rm{ssc}}$ does not appear to be in accord with the MFPT
either. These findings suggest that one simply estimates the
emission rate of neutrons over fission from the ratio $\Gamma_n/
\Gamma_{\rm{K}}$. Anything else does not seem to be in accord with
the assumption that fission can be described by the Kramers or
Smoluchowski equations, for the usually assumed form of the
potential. This does not rule out any effects related to a more
complicated dynamics, in particular if the initial stage of the
whole reaction is to be described with a different transport
theory. These findings may perhaps imply that some of the existing
statistical codes will have to be revised. The question of the
temperature dependence of nuclear transport does not seem to be a
closed one yet. A theoretical prediction has been given in
\cite{hiry}.

%Acknowledgements:
%This paper was partially supported by the ECT* ('STATE' contract)
%for a collaboration meeting on "Fission at finite thermal
%excitations" in April, 2002.

The authors benefitted greatly from a collaboration meeting on
"Fission at finite thermal excitations"  in April, 2002, sponsored
by the ECT* ('STATE' contract).  One of us (F.A.I.) would like to thank
the Physik Department of the TUM for the hospitality extended to him
during his stay at Garching.

\end{document}